\documentclass[12pt]{article}
\usepackage{amsbsy}
\usepackage{amsfonts}
\usepackage{amsmath}
\usepackage{amssymb}
\usepackage{apacite}
\usepackage{setspace}
\usepackage{graphicx}
\usepackage{bm,multicol}
\usepackage[english]{babel}
\usepackage{natbib}
\usepackage{hyperref}
\usepackage[T1]{fontenc}
\usepackage[utf8]{inputenc}
\usepackage{authblk}
\usepackage{xcolor}
\usepackage{booktabs}
\hypersetup{colorlinks=false,linkcolor=blue,urlcolor=blue}

\newcommand{\newc}{\newcommand}
\newc{\N}{\mbox{N}}
\newc{\1}{\bf{1}}

\begin{document}
\title{Comment on Vogels et al. (2024) ``Bayesian structure learning in undirected Gaussian graphical models'': An updated performance check using \texttt{BGGM}}

\author{Joris Mulder}


\thispagestyle{empty} \maketitle

\begin{abstract}
\noindent
\cite{vogels2024bayesian} presented an empirical comparison of Bayesian methods for
structure learning in undirected Gaussian graphical models. The method implemented in the R package
\texttt{BGGM} was run with equal prior probabilities for a null, negative,
and positive partial correlation, implying a prior inclusion
probability (equivalent to a prior graph density) of $\frac{2}{3}$. The other Bayesian methods in the comparison used a prior graph density of $0.2$ however. The data-generating densities ranged
from $0.01$ to $0.1$. This resulted in a substantial overestimation of the inclusion probabilities of absent edges by \texttt{BGGM}. For a fair comparison of the performance of the different methods, I repeated the simulation using
a prior inclusion probability of $0.2$ using \texttt{BGGM}. In this case, the performance of \texttt{BGGM} is comparable with the other Bayesian methods.
\end{abstract}


\section{Introduction}

There has been a considerable development of statistical approaches for
learning conditional dependency graphs. Such graphs are now used routinely in
psychology, where the nodes are observed variables such as symptoms or
questionnaire items and the edges are partial correlations
\citep{BorsboomCramer2013,EpskampBorsboomFried2018,EpskampFried2018}. The paper
by \cite{vogels2024bayesian}
presents a numerical simulation assessing the performance of a broad set of
Bayesian methods---among them the birth--death sampler of
\cite{MohammadiWit2015} implemented in \texttt{BDgraph}
\citep{MohammadiWit2019}, the stochastic search approach of \cite{Wang2015},
the graphical horseshoe of \cite{LiCraigBhadra2019}, the ECM algorithm of
\cite{GanNarisettyLiang2019}, the weighted proposal algorithm of
\cite{vandenBoom2022}, and the Bayes factor method in
\texttt{BGGM} \citep{WilliamsMulder2020JMP,WilliamsMulder2020JOSS}---together
with the non-Bayesian graphical lasso \citep{glasso}, for learning the
structure of a conditional dependence graph. Because of the breadth of available methods, such a comparison is useful.

The comparison is, however, sensitive to the prior settings used for each
method. Posterior model and
inclusion probabilities depend on the prior probability assigned to each model
\citep{ScottBerger2010}. In the case of \texttt{BGGM} the
setting that was used differed from that of the other Bayesian methods in a way
that materially affects two of the three performance criteria. The purpose of
this comment is to report the performance of \texttt{BGGM} under a prior
setting that is comparable to the one used for the other methods.

\section{Prior inclusion probability}

The performance of \texttt{BGGM} was assessed using estimates of the edge
inclusion probabilities obtained from
\[
P\big((i,j)\notin G \mid \mathbf{X}\big) =
\frac{BF_{0u}P(H_0)}{BF_{0u}P(H_0)+BF_{1u}P(H_1)+BF_{2u}P(H_2)},
\]
where $H_0:\rho_{ij}=0$, $H_1:\rho_{ij}>0$ and $H_2:\rho_{ij}<0$, and the
Bayes factors are obtained from a Savage--Dickey density ratio
\citep{Dickey1971,VerdinelliWasserman1995} applied to a normal approximation
of the posterior of the Fisher-transformed partial correlation
\citep{WilliamsMulder2020JMP}. In the performance check presented by \cite{vogels2024bayesian} equal prior probabilities were used for these three hypotheses,
$P(H_0)=P(H_1)=P(H_2)=1/3$, so that the prior probability that an edge is
present equals $2/3$. This differs from the prior used for the other Bayesian
methods in the comparison, which assumed a prior graph density of $0.2$
in their Equation (20), and it differed even more from the data-generating
densities, which ranged from $0.01$ to $0.1$.

Because the posterior inclusion probability is monotone in the prior odds,
assigning prior probability $2/3$ to the presence of an edge inflates the
inclusion probabilities of every edge, present or absent. This is visible in
the results: in Table 8 of \cite{vogels2024bayesian}, \texttt{BGGM} attains estimated inclusion probabilities of true absent edges as large as 
$Pr^-=0.88$ for $p=10$ and $n=20$, against $0.05$ to $0.18$ for all other
methods in the same cell. That outcome is a property of the prior setting rather
than of the method however. It should be noted that the ranking of edges, and
therefore the AUC, is unaffected by this choice, since a change in the prior
odds rescales the posterior odds by the same constant.
Note that the reported AUCs using \texttt{BGGM} by \cite{vogels2024bayesian} are also competitive with other Bayesian methods. For a fair comparison regarding the performance criterion $Pr^-$ (which will also affect the performance $Pr^+$), I repeat the simulation with \texttt{BGGM} using a prior inclusion
probability of $0.2$, matching the value used for the other Bayesian methods.

Note that the inclusion probabilities based on the exhaustive test of $H_0:\rho_{ij}=0$ vs $H_1:\rho_{ij}>0$ vs $H_2:\rho_{ij}<0$ are identical to those of a two-sided
test of $H_0:\rho_{ij}=0$ against $H_u:\rho_{ij}\neq 0$ using the same
$P(H_0)$: because the prior of $\rho_{ij}$ under $H_u$ is symmetric around
zero, the two one-sided hypotheses receive half of the remaining prior
probability each, which leaves $P(H_0\mid\mathbf{X})$ unchanged. For this reason, I used the default two-sided test of \texttt{BGGM} \citep{BGGM} to obtain the inclusion probabilities, which would be equivalent as when using the exhaustive test.

\section{Simulation}

I follow the design of Section 4.1 of \cite{vogels2024bayesian} exactly. Graphs are
generated with \texttt{BDgraph::bdgraph.sim}, in a Random and a Cluster
variant, and the precision matrix is drawn from $W_{G^*}(3,\mathbf{I})$\footnote{As the prior of \texttt{BDgraph} also follows a $G$-Wishart$(3,\mathbf{I})$ distribution, it exactly matches the data-generating mechanism of the precision matrices in the simulation. The performance reported for \texttt{BDgraph} is therefore maximized, which need not reflect its behaviour in general practice, where the prior will not match the data-generating mechanism. \texttt{BGGM} uses matrix $F$ prior \citep{MulderPericchi:2018} to construct stretched beta priors on $(-1,1)$ for the partial correlations.}; data
are then $n$ draws from $N(\mathbf{0},\mathbf{K}^{*-1})$. Performance is measured by the AUC of the edge
inclusion probabilities, and by $Pr^+$ and $Pr^-$, the mean inclusion
probability over the edges that are, respectively, present and absent in the
true graph. The reader is referred to \cite{vogels2024bayesian} for further details.

As \texttt{BGGM} was not
designed for $n<p$ problems, I restrict attention to the scenarios with $n>p$. This restriction is of limited
practical consequence for the applications \texttt{BGGM} was developed for: in
psychological network analysis the sample size is typically considerably
larger than the number of variables in the graph.

Two computational points are worth noting. First, the Gibbs sampler
underlying \texttt{BGGM} mixes slowly as $p$ grows, because the conditional
Wishart draws have degrees of freedom of order $p$. At $p=100$ I measured an
integrated autocorrelation time \citep{Geyer1992} of roughly $450$, using the
\texttt{effectiveSize} function of the \texttt{coda} package
\citep{Plummeretal2006}, so that the package default of $5000$ iterations
yields an effective sample size near $11$. I therefore
used $45{,}000$ post-burn-in iterations throughout, giving an effective sample
size near $100$ and a Monte Carlo error on the posterior standard deviation of
the Fisher-transformed partial correlations of about $7\%$. Second, the $p=1000$ scenarios were not attempted. Memory is not the obstacle:
the running posterior means and standard deviations of the partial
correlations are updated on the fly, so that the individual draws need not be
stored and memory use does not grow with the number of iterations. Computational time however poses a problem for such large neworks. At $p=1000$ a single iteration takes about $2.7$ seconds, and the
autocorrelation time grows with $p$, so that one fit with a comparable
effective sample size would require roughly two weeks. The analyses reported here use \texttt{BGGM} (CRAN version 2.2.0). The code is given in Appendix A.

\section{Results}

Tables \ref{tab:auc}, \ref{tab:prmin} and \ref{tab:prplus} report the AUC,
$Pr^-$ and $Pr^+$ respectively. To
allow the comparison across methods, each table reproduces the
corresponding table of \cite{vogels2024bayesian} in full, including the column
H-BGGM obtained there with a prior inclusion probability of $2/3$; the final column, labeled BGGM(0.2), gives my results under a prior inclusion probability of $0.2$.

\begin{table}[ht]
\noindent
\caption{AUC scores, averaged over 16 replications. The first eight numerical
columns are reproduced from Table 7 of \cite{vogels2024bayesian}. The results of H-BGGM were based on an inclusion probability of 2/3 using BGGM version 2.0.0. The last column, labeled BGGM(0.2), was obtained using BGGM version 2.2.0 and inclusion probability 0.2.}
\label{tab:auc}
\footnotesize
\setlength{\tabcolsep}{3.5pt}
\noindent\hspace*{-2cm}
\begin{tabular}{llrrrrrrrrrrr}
\toprule
 & & & & \multicolumn{8}{c}{\cite{vogels2024bayesian}} & \multicolumn{1}{c}{Updated} \\
\cmidrule(lr){5-12}\cmidrule(lr){13-13}
$p$ & Graph & Dens. & $n$ & glasso & RJ-WWA & BD-A & SS-O & ECM-B & G-MPLBD & K-HS & H-BGGM & BGGM(0.2)\\
\midrule
10  & Random  & 10\,\% & 20  & 0.70 & 0.72 & 0.73 & 0.73 & 0.70 & 0.72 & 0.70 & 0.67 & 0.69  \\
10  & Random  & 10\,\% & 350 & 0.94 & 0.95 & 0.95 & 0.93 & 0.91 & 0.95 & 0.94 & 0.93 & 0.94  \\
10  & Cluster & 10\,\% & 20  & 0.83 & 0.82 & 0.82 & 0.81 & 0.79 & 0.82 & 0.81 & 0.75 & 0.72  \\
10  & Cluster & 10\,\% & 350 & 0.93 & 0.93 & 0.93 & 0.91 & 0.92 & 0.94 & 0.92 & 0.92 & 0.92  \\
100 & Random  & 1\,\%  & 700 & 0.83 & 0.96 & 0.96 & 0.96 & 0.94 & 0.95 & 0.96 & 0.95 & 0.95  \\
100 & Random  & 10\,\% & 700 & 0.78 & 0.92 & 0.91 & 0.89 & 0.80 & 0.90 & 0.92 & 0.87 & 0.87  \\
100 & Cluster & 1\,\%  & 700 & 0.95 & 0.96 & 0.96 & 0.94 & 0.94 & 0.94 & 0.95 & 0.94 & 0.94  \\
100 & Cluster & 10\,\% & 700 & 0.84 & 0.94 & 0.93 & 0.91 & 0.84 & 0.92 & 0.93 & 0.87 & 0.87  \\
\bottomrule
\end{tabular}
\end{table}

\begin{table}[ht]
\noindent
\caption{$Pr^-$, the mean inclusion probability of absent edges from the true
graph; lower is better. Columns as in Table \ref{tab:auc}, with the first eight
numerical columns reproduced from Table 8 of \cite{vogels2024bayesian}. The results of H-BGGM were based on an inclusion probability of 2/3 using BGGM version 2.0.0. The last column, labeled BGGM(0.2), was obtained using BGGM version 2.2.0 and inclusion probability 0.2.}
\label{tab:prmin}
\footnotesize
\setlength{\tabcolsep}{3.5pt}
\noindent\hspace*{-2cm}
\begin{tabular}{llrrrrrrrrrrr}
\toprule
 & & & & \multicolumn{8}{c}{\cite{vogels2024bayesian}} & \multicolumn{1}{c}{Updated} \\
\cmidrule(lr){5-12}\cmidrule(lr){13-13}
$p$ & Graph & Dens. & $n$ & glasso & RJ-WWA & BD-A & SS-O & ECM-B & G-MPLBD & K-HS & H-BGGM & BGGM(0.2)\\
\midrule
10  & Random  & 10\,\% & 20  & 0.05 & 0.09 & 0.10 & 0.08 & 0.18 & 0.08 & 0.06 & 0.88 & 0.15  \\
10  & Random  & 10\,\% & 350 & 0.05 & 0.02 & 0.03 & 0.03 & 0.15 & 0.01 & 0.09 & 0.33 & 0.05  \\
10  & Cluster & 10\,\% & 20  & 0.04 & 0.08 & 0.09 & 0.08 & 0.20 & 0.07 & 0.02 & 0.88 & 0.15  \\
10  & Cluster & 10\,\% & 350 & 0.06 & 0.02 & 0.03 & 0.03 & 0.12 & 0.01 & 0.07 & 0.31 & 0.04  \\
100 & Random  & 1\,\%  & 700 & 0.01 & 0.01 & 0.02 & 0.03 & 0.11 & 0.01 & 0.01 & 0.23 & 0.03  \\
100 & Random  & 10\,\% & 700 & 0.17 & 0.03 & 0.04 & 0.05 & 0.15 & 0.00 & 0.04 & 0.23 & 0.03  \\
100 & Cluster & 1\,\%  & 700 & 0.00 & 0.01 & 0.02 & 0.03 & 0.14 & 0.01 & 0.00 & 0.23 & 0.03  \\
100 & Cluster & 10\,\% & 700 & 0.11 & 0.02 & 0.03 & 0.05 & 0.12 & 0.00 & 0.03 & 0.23 & 0.03  \\
\bottomrule
\end{tabular}
\end{table}

\begin{table}[ht]
\noindent
\caption{$Pr^+$, the mean inclusion probability of present edges in the true
graph; higher is better. Columns as in Table \ref{tab:auc}, with the first
eight numerical columns reproduced from Table 9 of \cite{vogels2024bayesian}. The results of H-BGGM were based on an inclusion probability of 2/3 using BGGM version 2.0.0. The last column, labeled BGGM(0.2), was obtained using BGGM version 2.2.0 and inclusion probability 0.2.}
\label{tab:prplus}
\footnotesize
\setlength{\tabcolsep}{3.5pt}
\noindent\hspace*{-2cm}
\begin{tabular}{llrrrrrrrrrrr}
\toprule
 & & & & \multicolumn{8}{c}{\cite{vogels2024bayesian}} & \multicolumn{1}{c}{Updated} \\
\cmidrule(lr){5-12}\cmidrule(lr){13-13}
$p$ & Graph & Dens. & $n$ & glasso & RJ-WWA & BD-A & SS-O & ECM-B & G-MPLBD & K-HS & H-BGGM & BGGM(0.2) \\
\midrule
10  & Random  & 10\,\% & 20  & 0.38 & 0.38 & 0.40 & 0.35 & 0.31 & 0.40 & 0.39 & 0.90 & 0.30 \\
10  & Random  & 10\,\% & 350 & 0.87 & 0.77 & 0.78 & 0.56 & 0.69 & 0.81 & 0.85 & 0.90 & 0.78 \\
10  & Cluster & 10\,\% & 20  & 0.45 & 0.45 & 0.47 & 0.40 & 0.35 & 0.50 & 0.45 & 0.91 & 0.39 \\
10  & Cluster & 10\,\% & 350 & 0.87 & 0.79 & 0.80 & 0.58 & 0.73 & 0.82 & 0.88 & 0.89 & 0.76 \\
100 & Random  & 1\,\%  & 700 & 0.83 & 0.84 & 0.84 & 0.63 & 0.82 & 0.84 & 0.87 & 0.90 & 0.80 \\
100 & Random  & 10\,\% & 700 & 0.59 & 0.73 & 0.72 & 0.57 & 0.58 & 0.68 & 0.79 & 0.76 & 0.55 \\
100 & Cluster & 1\,\%  & 700 & 0.81 & 0.83 & 0.84 & 0.61 & 0.83 & 0.84 & 0.81 & 0.90 & 0.80 \\
100 & Cluster & 10\,\% & 700 & 0.59 & 0.73 & 0.73 & 0.59 & 0.63 & 0.68 & 0.80 & 0.77 & 0.71 \\
\bottomrule
\end{tabular}
\end{table}

\paragraph{AUC.} The AUC values are unchanged from those reported by
\cite{vogels2024bayesian} taking the Monte Carlo error of approximately $0.025$ into account. This equivalence is as expected. A change in the prior inclusion probability
multiplies every Bayes factor by the same constant and therefore leaves the
ranking of the edges, and hence the AUC, intact. Relative to the other methods, the AUC of \texttt{BGGM} is roughly comparable.

\paragraph{$Pr^-$.} This is where the prior setting affects the results. Under a
prior inclusion probability of $0.2$, $Pr^-$ falls from $0.88$ to $0.15$ for
$p=10$ and $n=20$, and from $0.23$ to $0.03$ for $p=100$ and $n=700$, so that
\texttt{BGGM} is no longer an outlier but lies within the spread of the other
methods. In the smallest scenarios ($n=20$) a $Pr^-$ of $0.15$ is still at the
high end of the field ($0.02$ to $0.20$), which is to be expected: with this
little information the inclusion probabilities remain close to the prior value
of $0.2$, whereas the regularized methods shrink them further towards zero. For
the larger sample sizes, which are more representative of applications in
psychology, $Pr^-$ lies between $0.03$ and $0.05$, comparable to or better than
most of the other methods.

\paragraph{$Pr^+$.} The reduction in $Pr^-$ is accompanied, as it must be, by
a reduction in $Pr^+$: fewer edges are declared present overall. For $p=10$
and $n=20$, $Pr^+$ falls from $0.90$ to $0.30$, which places \texttt{BGGM}
alongside SS-O ($0.35$) and ECM-BAGUS ($0.31$) rather than far above every
method in the row; for $p=100$, $n=700$ at $1\%$ density it falls from $0.90$
to $0.80$, against $0.63$ to $0.87$ for the others. Read together with $Pr^-$,
the two criteria show that the original setting placed \texttt{BGGM} at a very
different point on the inclusion scale than the other methods rather than at a
different level of accuracy, and that $Pr^-$ and $Pr^+$ should not be
interpreted separately.


\section{Conclusion}

The aim of this note was to present an updated performance check of \texttt{BGGM} using a prior setting that is comparable with the employed priors of the other Bayesian methods. The results show that the performance of
\texttt{BGGM} on $Pr^-$ and $Pr^+$ falls within the spread of the other
Bayesian methods. The updated results show that
the original results of \cite{vogels2024bayesian} severely overstated the difference by a wide margin, and for a
reason that has nothing to do with the method.


More generally, these results illustrate that comparisons of this kind are
informative only to the extent that the prior settings of the competing
methods are placed on a common footing:
the prior on the graph density, as well as the prior on the magnitude of the
nonzero edges, which was touched on here only in a footnote, both affect
the behavior of these methods. Therefore it is generally recommendable to be explicit about the employed prior settings of different methods in empirical performance checks.

\bibliographystyle{apacite}
\bibliography{refs_comment}

\appendix

\section{R code for BGGM analysis}

The full simulation script is available at
\url{https://github.com/jomulder/BGGM}. The core of the analysis for a single
data set is as follows.

\begin{verbatim}
library(BDgraph); library(BGGM)

p <- 100; n <- 700; graph <- "random"; prob <- 0.1
q <- 0.2                       # prior inclusion probability

set.seed(1)
sim <- bdgraph.sim(p = p, n = n, graph = graph, prob = prob,
                   size = NULL, type = "Gaussian", vis = FALSE)
response <- sim$G[upper.tri(sim$G)]

fit <- explore(sim$data, iter = 45000, prior_sd = 0.5, seed = 1001,
               progress = FALSE, store_post_draws = FALSE)

sel <- BGGM:::select.explore(fit, alternative = "two.sided",
                             prior.prob.H0 = 1 - q)
pip <- sel$incl_prob[upper.tri(sel$incl_prob)]

rk  <- rank(pip)
n1  <- sum(response); n0 <- length(response) - n1
AUC <- (sum(rk[response == 1]) - n1 * (n1 + 1) / 2) / (n1 * n0)
c(AUC = AUC, Prplus = mean(pip[response == 1]),
  Prmin = mean(pip[response == 0]))
\end{verbatim}

\end{document}